\documentclass[sigconf]{acmart}

\usepackage{multirow}
\usepackage{booktabs}   

\usepackage{pifont}

\usepackage{balance}

\newcommand{\bodyrefbreak}{%
	\balance
	\clearpage
}

\AtBeginDocument{%
	}

\setcopyright{acmlicensed}

\copyrightyear{2026}
\acmYear{2026}
\setcopyright{cc}
\setcctype{by}
\acmConference[MM '26]
{Proceedings of the 35th ACM International Conference on Multimedia}
{November 10--14, 2026}
{Rio de Janeiro, Brazil.}
\acmBooktitle{Proceedings of the 35th ACM International Conference on Multimedia
	(MM '26), November 10--14, 2026, Rio de Janeiro, Brazil}
\acmISBN{979-8-4007-2213-4/2026/11}
\acmDOI{10.1145/3767308.3836537}
\begin{document}

\title{ParaJSCC: A Parameterized Framework for Reusable Multimodal Joint Source-Channel Coding}

	\author{Kemi Chen}
\affiliation{%
	\institution{Fuzhou University}
	\city{Fuzhou}
	\country{China}
}
\email{arice\_chen@163.com}

\author{Mingkai Chen}
\affiliation{%
	\institution{Nanjing University of Posts and Telecommunications}
	\city{Nanjing}
	\state{Jiangsu}
	\country{China}
}
\email{mkchen@njupt.edu.cn}

\author{Youjia Chen}
\affiliation{%
	\institution{Fuzhou University}
	\city{Fuzhou}
	\country{China}
}
\email{youjia.chen@fzu.edu.cn}

\author{Qian Liu}
\affiliation{%
	\institution{Dalian University of Technology}
	\city{Dalian}
	\state{Liaoning}
	\country{China}
}
\email{qianliu@dlut.edu.cn}

\author{Wei Gao}
\affiliation{%
	\institution{Peking University}
	\city{Beijing}
	\country{China}
}
\email{gaowei262@pku.edu.cn}

\author{Tiesong Zhao}
\authornote{This work is supported by National Natural Science Foundation of China (Grant No.
	62571131) and Funds for the Innovation of Policing Science and Technology, Fujian province (Grant number: 2025Y0070). Corresponding author: Tiesong Zhao.}
\affiliation{%
	\institution{Fuzhou University}
	\city{Fuzhou}
	\country{China}
}
\email{t.zhao@fzu.edu.cn}
\renewcommand{\shortauthors}{Kemi Chen et al.}

\begin{abstract}
  Multimodal signals, such as visual, audio, and tactile data, are increasingly maintained as persistent digital assets in immersive communication systems and digital twins. In these settings, the same multimodal content is repeatedly accessed by heterogeneous receivers with varying modality and bandwidth requirements. Existing compression and Joint Source-Channel Coding (JSCC) methods typically follow a per-request encoding paradigm, resulting in redundant computation and low efficiency during repeated access. To address this issue, we propose ParaJSCC, a multimodal JSCC framework designed for reusable representation serving. ParaJSCC converts each multimodal sample offline at the cloud/content server into a compact, quantized parameter package, which is then stored at the edge serving node for low-latency access. During serving, only the subset required by the current request is transmitted over the wireless channel, followed by lightweight decoding at the receiver. The framework employs a progressive shared-private parameterization to support modality-selective transmission and scalable reconstruction under varying bandwidth constraints. Experiments on multimodal datasets show that ParaJSCC significantly reduces online latency (e.g., from 17.18~ms to 4.34~ms for image-only requests and from 43.96~ms to 11.21~ms for full multimodal requests) and transmission rate (by 47.8\%--51.2\% for selective requests), while maintaining strong reconstruction quality under noisy channels.
\end{abstract}



\maketitle

\section{Introduction}
\begin{figure}[H]
	\centering
	\includegraphics[width=0.9\linewidth]{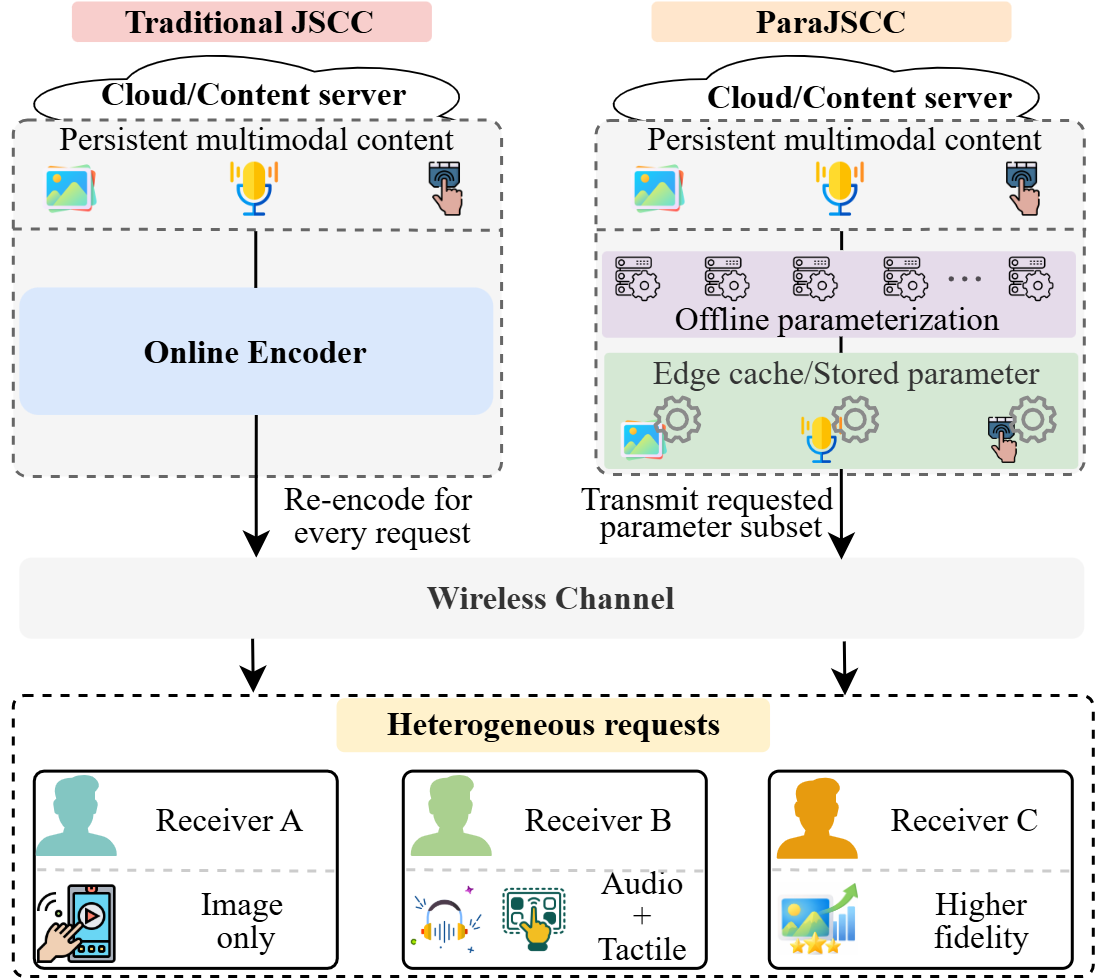}
	\caption{Application scenario of repeated multimodal serving. Traditional JSCC re-encodes the raw multimodal source for each request, whereas ParaJSCC converts persistent content into a reusable parameter package stored at the edge. For each heterogeneous request, only the required parameter subset is transmitted for reconstruction.}
	\label{fig:parajscc}
\end{figure}

Multimodal content is increasingly stored as persistent digital assets rather than transmitted only once. Examples include multimodal environment records, digital twins, and long-term sensing archives that continuously accumulate heterogeneous data and support diverse applications~\cite{tao2024advancements,TUM2TWIN2026}. In such settings, the same multimodal content may be accessed repeatedly by different users or applications with diverse requirements~\cite{zenseact2023}. This trend raises a new challenge: how to efficiently serve persistent multimodal content under repeated and heterogeneous access patterns.

We consider a repeated-access multimodal serving system with a cloud-edge-receiver pipeline. Raw content is stored on a cloud server, where each sample is optimized offline and converted into a quantized parameter package. The package is then deployed to an edge node for low-latency storage and online access. During serving, receivers request a subset of modalities and a progressive level, and the edge transmits only the corresponding subset of parameters for reconstruction.

Existing communication systems are mainly designed for request-time delivery rather than reusable serving. Traditional compression pipelines and neural Joint Source-Channel Coding (JSCC) methods typically follow a per-request encoding paradigm, where the raw source is re-encoded for every transmission~\cite{DeepJSCC2019,OFDM2021,MambaJSCC2025}. Although effective for one-time delivery, this paradigm becomes inefficient under repeated access because it incurs redundant computation and reduces online responsiveness. As illustrated in Fig.~\ref{fig:parajscc}, the same source must be processed repeatedly for different requests.

Recent learned communication methods have substantially improved robust transmission for both individual modalities and unified multimodal settings, including image JSCC~\cite{zhang2022toward,lyu2024semantic,huang2026robust}, semantic speech communication~\cite{weng2025robust,tian2026large,han2026error}, and vibrotactile coding~\cite{zeng2020perception,ozawa2025multi,nockenberg2025deep}. These methods show strong robustness under noisy and bandwidth-limited conditions. However, they still operate in a request-time manner: they do not support cross-request reuse or request-aware subset transmission from a pre-prepared asset. Meanwhile, Implicit Neural Representation (INR)-based compression and communication methods introduce instance-specific parameterized representations, but they mainly focus on compact representation and direct reconstruction of individual samples rather than structured serving for repeated multimodal access.

To address this gap, we propose ParaJSCC, a framework that shifts communication from request-time source encoding to reusable representation serving. Each multimodal sample is converted offline into a compact parameter package and deployed at the edge for future access. During online serving, only the subset required by the current request is transmitted to the receiver, followed by lightweight decoding. Accordingly, ParaJSCC is better viewed as a serving-oriented communication framework than as a conventional instance-specific JSCC model.

To support heterogeneous requests, ParaJSCC adopts a progressive shared-private parameterization. Shared parameters capture cross-modal structure, while modality-specific parameters encode unique details. This design enables modality-selective transmission and coarse-to-fine reconstruction under varying bandwidth constraints. Unlike INR or Implicit-JSCC methods, which rely on largely flat instance representations, and unlike unified multimodal communication methods such as U-DeepSC, which still require online encoding for each request, ParaJSCC integrates cross-modal sharing, request-aware transmission, and progressive reconstruction into a reusable serving framework.

The main contributions of this work are summarized as follows:

\begin{itemize}
	\item \textbf{Problem formulation.} We introduce repeated-access multimodal communication, a setting in which persistent multimodal content is delivered to heterogeneous receivers with diverse modality and fidelity requirements.
	
	\item \textbf{Framework design.} We propose ParaJSCC, a communication framework that decouples offline representation preparation from online request-aware transmission, enabling efficient reuse across repeated accesses.
	
	\item \textbf{Representation modeling.} We develop a progressive shared-private parameterization that supports cross-modal information sharing, selective modality transmission, and scalable reconstruction within a unified structure.
	
	\item \textbf{System validation.} We implement a concrete network instantiation of ParaJSCC with shared-to-private decoding and cross-modal alignment, and demonstrate its effectiveness in multimodal communication tasks involving images, audio, and vibrotactile signals.
\end{itemize}

\begin{figure*}[t]
	\centering
	\includegraphics[width=0.9\linewidth]{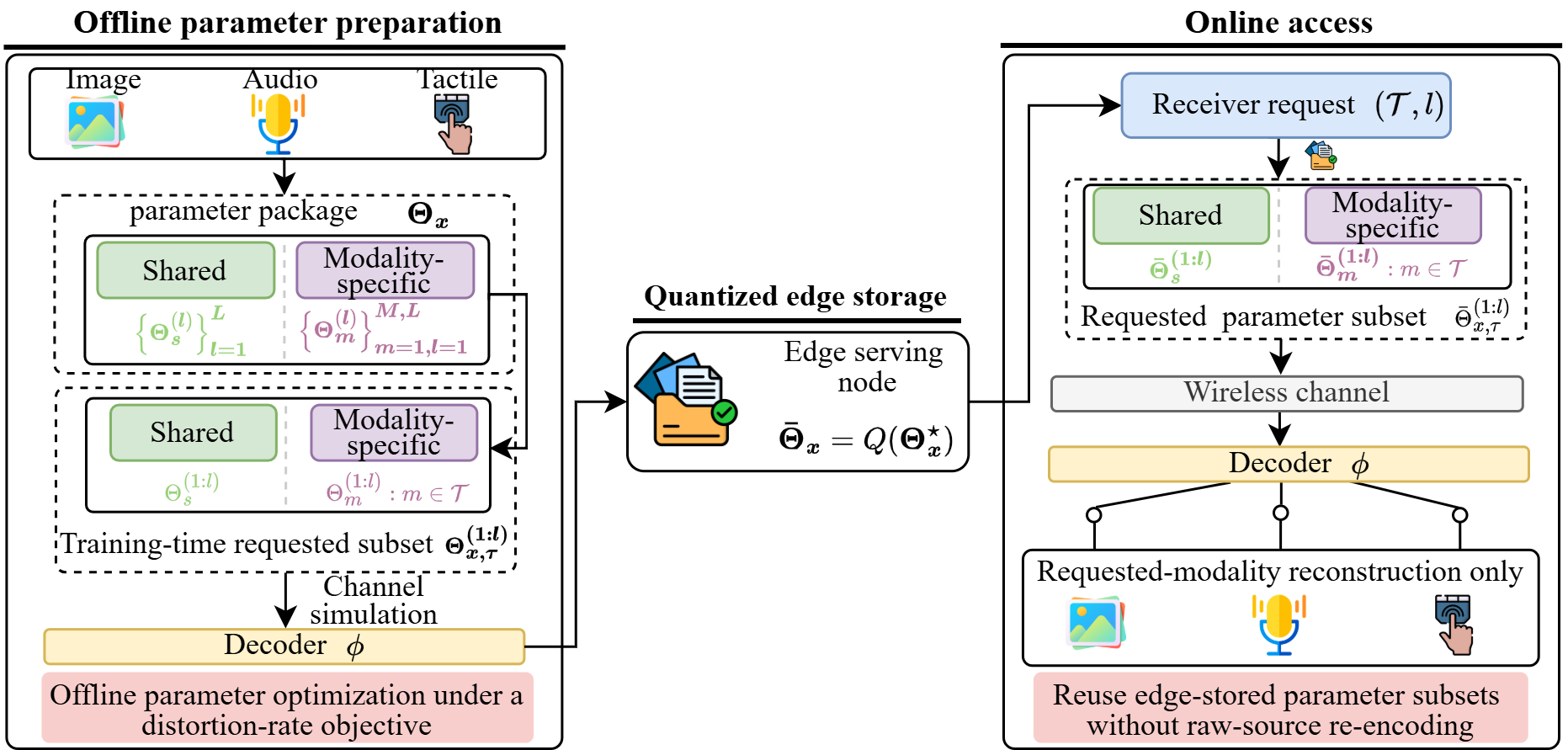}
	\caption{Overall pipeline of ParaJSCC. Each multimodal sample is converted offline into a quantized shared-private parameter package stored at the edge. During training, sampled requests guide subset selection for channel simulation and distortion-rate optimization. During online access, only the requested subset is transmitted for reconstruction.}
	\label{fig:JSCC_vs_ParaJSCC}
\end{figure*}

\section{Related Work}

\subsection{Deep JSCC}

JSCC has become an effective alternative to separation-based communication, especially under noisy and bandwidth-constrained wireless conditions. Deep JSCC has achieved strong performance in image transmission~\cite{ADJSCC2022,JSCCformer2024,CDDM2024,Swinjscc2025}, while related advances in speech and semantic communication, such as DeepSC-S~\cite{DeepSC-S2021} and DNN-based approaches~\cite{DNN2023,DNN22025}, further demonstrate the benefits of jointly optimizing signal delivery and recovery. Compared with image and speech communication, vibrotactile communication remains less explored. Existing studies mainly focus on source coding and perceptual compression, including multimedia-inspired codecs~\cite{zeng2020perception}, multi-channel schemes based on clustering and differential coding~\cite{MVibCode2023}, and perceptually motivated methods such as PVC-SLP~\cite{PVC-SLP2020} and VC-PWQ~\cite{VC-PWQ2021}. More recent learning-based codecs, including RNVC~\cite{RNVC2022} and VTSC~\cite{VTSC2025}, further improve coding efficiency and rate--distortion performance. However, only limited work has investigated JSCC-based vibrotactile transmission~\cite{ogura2023semantic,ogura2025joint}.

Overall, existing JSCC and modality-specific communication methods mainly target point-to-point transmission or single-modality optimization. They do not address repeated multimodal serving under heterogeneous requests, nor do they provide reusable representations for request-aware subset delivery. In contrast, our work focuses on repeated-access multimodal communication, where offline-prepared representations can be reused across diverse requests.

\subsection{Multimodal Communication}

Recent studies have extended semantic communication to multimodal settings~\cite{liu2024cross}. U-DeepSC~\cite{U-DeepSC2024} provides a unified framework for images, text, and speech, while SyncSC~\cite{PacSC2025} studies synchronous multimodal transmission through semantic and temporal alignment. Cross-modal correlation has also been explored in tactile and haptic communication; for example, cross-modal haptic compression~\cite{lu2025cross} uses visual semantics to improve tactile coding efficiency.

Despite these advances, existing multimodal communication and compression methods are still mainly designed for per-transmission delivery. Unified multimodal frameworks support multiple modalities within a single model, and cross-modal compression methods exploit inter-modal correlation for higher efficiency. However, they still rely on online processing for each request and do not support cross-request reuse or request-aware subset transmission from a reusable representation. In contrast, our work targets repeated-access multimodal communication with offline-prepared representations reusable across heterogeneous requests.

\subsection{INR-Based Compression and Communication}

INR-based compression represents individual source instances with lightweight neural functions. Early work such as COIN++~\cite{coin++2022} shows that instance-specific representations can be learned by modulating a shared base network with quantization and entropy coding. Subsequent methods, including COOL-CHIC~\cite{ladune2023cool}, C3~\cite{kim2024c3}, and LotteryCodec~\cite{LotteryCodec2025}, further improve rate--distortion performance and efficiency. This idea has also been extended to communication. CSI-INR~\cite{wu2025mimo} models channel state information as neural functions, while Implicit-JSCC~\cite{Implicit-JSCC2025} introduces instance-specific optimization into JSCC, showing that sample-adaptive representations can replace universal encoder--decoder models.

However, existing INR-based methods mainly target compact representation and direct reconstruction of individual samples, typically in one-shot or single-modality settings. Even Implicit-JSCC-style approaches treat the optimized representation as a largely flat instance-level structure, without supporting repeated multimodal serving, request-aware subset transmission, or progressive reconstruction. Therefore, their main limitation is not the absence of instance-specific parameterization, but the lack of a serving-oriented formulation. In contrast, our method organizes instance-specific representations into a progressive shared-private package for repeated multimodal requests, enabling reusable serving with request-aware subset transmission.

\section{Methodology}
\label{Methodology}

We formulate repeated-access multimodal communication as a reusable parameter serving problem in a cloud-edge-receiver pipeline (Fig.~\ref{fig:JSCC_vs_ParaJSCC}). Instead of encoding the raw multimodal source for each request, ParaJSCC performs one-time offline optimization at the cloud/content server and converts each sample into a compact parameter package for edge storage. During online serving, the edge retrieves and transmits only the parameter subset required by the current request. This formulation decouples offline representation preparation from online request-aware transmission.

\subsection{Problem Setup}

Consider a multimodal source \(\boldsymbol{X}=\{\boldsymbol{X}_1,\dots,\boldsymbol{X}_M\}\), where \(\boldsymbol{X}_m\) denotes modality \(m\), and let \(\boldsymbol{x}=\{\boldsymbol{x}_1,\dots,\boldsymbol{x}_M\}\) denote one realization. A receiver request is specified by a modality subset \(\mathcal{T}\subseteq\{1,\dots,M\}\) and a progressive level \(l\in\{1,\dots,L\}\), where \(L\) is the total number of progressive levels. To model heterogeneous access patterns, we define a request distribution \(p(\mathcal{T},l)\).

Given a request \((\mathcal{T},l)\), the receiver reconstructs
\begin{equation}
	\hat{\boldsymbol{x}}_{\mathcal{T}}^{(l)}=\{\hat{\boldsymbol{x}}_m^{(l)}\}_{m\in\mathcal{T}},
\end{equation}
using the transmitted parameter groups from levels \(1{:}l\), with the objective of minimizing reconstruction distortion under communication constraints.

For each sample \(\boldsymbol{x}\), ParaJSCC builds a sample-specific parameter package
\begin{equation}
	\Theta_{\boldsymbol{x}} =
	\left\{
	\{\boldsymbol{\Theta}_s^{(l)}\}_{l=1}^{L},
	\{\boldsymbol{\Theta}_m^{(l)} \mid m=1,\dots,M,\; l=1,\dots,L\}
	\right\},
\end{equation}
where \(\boldsymbol{\Theta}_s^{(l)} \in \mathbb{R}^{d_s^{(l)}}\) denotes the shared parameter group at level \(l\), and \(\boldsymbol{\Theta}_m^{(l)} \in \mathbb{R}^{d_m^{(l)}}\) denotes the modality-specific parameter group for modality \(m\). The representation combines shared-private decomposition across modalities with progressive refinement across levels. As a result, cross-modal common structure and modality-specific details are organized separately and can be transmitted incrementally according to the request.

For a request \((\mathcal{T},l)\), the transmitted subset is
\begin{equation}
	\Theta_{\boldsymbol{x},\mathcal{T}}^{(1:l)}
	=
	\Big\{
	\boldsymbol{\Theta}_s^{(1:l)},
	\boldsymbol{\Theta}_m^{(1:l)} : m \in \mathcal{T}
	\Big\},
\end{equation}
where \(\boldsymbol{\Theta}_s^{(1:l)}=\{\boldsymbol{\Theta}_s^{(j)}\}_{j=1}^{l}\) and
\(\boldsymbol{\Theta}_m^{(1:l)}=\{\boldsymbol{\Theta}_m^{(j)}\}_{j=1}^{l}\). Therefore, ParaJSCC serves heterogeneous requests by transmitting all shared parameters up to the requested level, together with only the modality-specific parameters associated with the requested modalities.

\subsection{Offline Objective}

For each sample \(\boldsymbol{x}\), the package \(\Theta_{\boldsymbol{x}}\) is optimized offline by minimizing the expected reconstruction distortion over varying requests and channel conditions, while penalizing transmission cost. Let \(d_m(\cdot,\cdot)\) denote the distortion metric for modality \(m\), with weight \(\lambda_m\). For a requested subset \(\mathcal{T}\), we define the request-level distortion as
\begin{equation}
	d_{\mathcal{T}}(\boldsymbol{x},\hat{\boldsymbol{x}}_{\mathcal{T}})
	=
	\sum_{m \in \mathcal{T}}
	\lambda_m d_m(\boldsymbol{x}_m,\hat{\boldsymbol{x}}_m),
	\label{rm}
\end{equation}
which measures the weighted reconstruction distortion over the requested modalities.

Beyond accurate reconstruction, we also encourage the shared parameters to capture cross-modal common structure. To this end, we introduce a shared-only embedding alignment regularizer. Let \(\boldsymbol{f}_s^{(l)}\) denote the shared-only decoded representation obtained from \(\boldsymbol{\Theta}_s^{(1:l)}\), and let \(\boldsymbol{e}_m^{(s,l)}\in\mathbb{R}^D\) denote the embedding obtained by projecting \(\boldsymbol{f}_s^{(l)}\) into the space of modality \(m\). Define
\begin{equation}
	\mathcal{P}(\mathcal{T})=\{(m,n): m,n\in\mathcal{T},\, m<n\}.
\end{equation}
The cross-modal regularization term is
\begin{equation}
	\small
	\mathcal{L}_{\mathrm{cm}}^{(l)}(\boldsymbol{x},\mathcal{T})
	=
	\begin{cases}
		\displaystyle
		\frac{1}{|\mathcal{P}(\mathcal{T})|}
		\sum_{(m,n)\in\mathcal{P}(\mathcal{T})}
		\left\| \boldsymbol{e}_m^{(s,l)} - \boldsymbol{e}_n^{(s,l)} \right\|_2^2,
		& \text{if } |\mathcal{T}| \ge 2, \\[0.5ex]
		0,
		& \text{if } |\mathcal{T}| < 2.
	\end{cases}
\end{equation}

During offline optimization, transmission is simulated by differentiable quantization, channel mapping, and decoding:
\begin{equation}
	\left\{
	\begin{aligned}
		\boldsymbol{s}_{\boldsymbol{x},\mathcal{T}}^{(1:l)}
		&=
		g\!\left(
		\widetilde{Q}\!\left(\Theta_{\boldsymbol{x},\mathcal{T}}^{(1:l)}\right)
		\right),\\
		\boldsymbol{y}_{\boldsymbol{x},\mathcal{T}}^{(1:l)}
		&=
		\mathcal{C}\!\left(
		\boldsymbol{s}_{\boldsymbol{x},\mathcal{T}}^{(1:l)};\xi
		\right),\\
		\hat{\boldsymbol{x}}_{\mathcal{T}}^{(l)}
		&=
		D_\phi\!\left(
		\boldsymbol{y}_{\boldsymbol{x},\mathcal{T}}^{(1:l)}
		\right),
	\end{aligned}
	\right.
\end{equation}
where \(\widetilde{Q}(\cdot)\) is a differentiable approximation to quantization, \(g(\cdot)\) is a fixed mapping from quantized parameters to channel symbols, and \(D_\phi(\cdot)\) is a decoder with globally shared parameters \(\phi\).

Let
\begin{equation}
	B_{\boldsymbol{x}} \triangleq \sum_{m=1}^{M} |\boldsymbol{x}_m|
\end{equation}
denote the total number of real-valued source symbols in sample \(\boldsymbol{x}\). For a request \((\mathcal{T},l)\), the transmitted subset contains all shared parameters up to level \(l\) and the modality-specific parameters of the requested modalities up to level \(l\). Let \(\kappa_s^{(j)}\) and \(\kappa_m^{(j)}\) denote the transmission expansion factors associated with \(\boldsymbol{\Theta}_s^{(j)}\) and \(\boldsymbol{\Theta}_m^{(j)}\), respectively. The resulting real-valued transmission cost is
\begin{equation}
	K_{\boldsymbol{x},\mathcal{T}}^{(l)}
	=
	\sum_{j=1}^{l}\kappa_s^{(j)} d_s^{(j)}
	+
	\sum_{m\in\mathcal{T}}\sum_{j=1}^{l}\kappa_m^{(j)} d_m^{(j)}.
\end{equation}

Assuming a complex-valued channel, where two real dimensions correspond to one channel use, the request-aware bandwidth ratio is
\begin{equation}
	R_{\boldsymbol{x},\mathcal{T}}^{(l)}
	=
	\frac{K_{\boldsymbol{x},\mathcal{T}}^{(l)}}{2B_{\boldsymbol{x}}}
	=
	R_{s,\boldsymbol{x}}^{(l)} + R_{p,\boldsymbol{x},\mathcal{T}}^{(l)},
\end{equation}
where
\begin{equation}
	R_{s,\boldsymbol{x}}^{(l)}
	=
	\frac{\sum_{j=1}^{l}\kappa_s^{(j)} d_s^{(j)}}{2B_{\boldsymbol{x}}},
\end{equation}
and
\begin{equation}
	R_{p,\boldsymbol{x},\mathcal{T}}^{(l)}
	=
	\frac{\sum_{m\in\mathcal{T}}\sum_{j=1}^{l}\kappa_m^{(j)} d_m^{(j)}}{2B_{\boldsymbol{x}}}.
\end{equation}

The resulting offline objective is
\begin{equation}
	\small
	\mathcal{J}_{\boldsymbol{x}}(\Theta_{\boldsymbol{x}})
	=
	\mathbb{E}_{(\mathcal{T},\,l)\sim p(\mathcal{T},l),\,\xi}
	\left[
	d_{\mathcal{T}}\!\left(\boldsymbol{x},\hat{\boldsymbol{x}}_{\mathcal{T}}^{(l)}\right)
	+
	\beta R_{\boldsymbol{x},\mathcal{T}}^{(l)}
	+
	\gamma \mathcal{L}_{\mathrm{cm}}^{(l)}(\boldsymbol{x},\mathcal{T})
	\right],
	\label{batagama}
\end{equation}
where \(\beta \ge 0\) controls the distortion--rate trade-off and \(\gamma \ge 0\) controls the strength of cross-modal regularization. The optimized package is then given by
\begin{equation}
	\Theta_{\boldsymbol{x}}^\star
	=
	\arg\min_{\Theta_{\boldsymbol{x}}}\; \mathcal{J}_{\boldsymbol{x}}(\Theta_{\boldsymbol{x}}).
\end{equation}

\subsection{Quantized Storage and Online Transmission}

After offline optimization at the cloud/content server, the optimized package is quantized and stored at the edge serving node:
\begin{equation}
	\bar{\Theta}_{\boldsymbol{x}} = Q(\Theta_{\boldsymbol{x}}^\star).
\end{equation}
Each shared or modality-specific tensor is quantized separately at each progressive level using a unified bit-width. The stored package contains the quantized values together with lightweight metadata for efficient subset retrieval. Unless otherwise specified, the reported transmission rate accounts only for the subset transmitted for the current request, rather than the storage cost of the full package prepared offline.

Given a request \((\mathcal{T},l)\), the edge serving node retrieves the corresponding subset
\begin{equation}
	\bar{\Theta}_{\boldsymbol{x},\mathcal{T}}^{(1:l)}
	=
	\Big\{
	\bar{\boldsymbol{\Theta}}_s^{(1:l)},
	\bar{\boldsymbol{\Theta}}_m^{(1:l)} : m \in \mathcal{T}
	\Big\},
\end{equation}
maps it to channel symbols, and transmits it through the channel:
\begin{equation}
	\boldsymbol{s}_{\boldsymbol{x},\mathcal{T}}^{(1:l)} = g\!\left(\bar{\Theta}_{\boldsymbol{x},\mathcal{T}}^{(1:l)}\right), \qquad
	\boldsymbol{y}_{\boldsymbol{x},\mathcal{T}}^{(1:l)} = \mathcal{C}\big(\boldsymbol{s}_{\boldsymbol{x},\mathcal{T}}^{(1:l)};\xi\big).
\end{equation}
The receiver then reconstructs the requested modalities as
\begin{equation}
	\hat{\boldsymbol{x}}_{\mathcal{T}}^{(l)} = D_\phi\big(\boldsymbol{y}_{\boldsymbol{x},\mathcal{T}}^{(1:l)}\big).
\end{equation}

\subsection{Break-Even Access Analysis}

Because ParaJSCC moves computation from per-request online encoding to one-time offline preparation, its practical advantage depends on the access frequency of a deployed multimodal asset. We define the break-even access count \(N^\star\) as the minimum integer number of accesses at which the cumulative latency of ParaJSCC does not exceed that of a baseline per-request encoding system:
\begin{equation}
	{\small
		N^\star=
		\left\lceil
		\frac{T_{\mathrm{offline}}^{\mathrm{Para}}}
		{\left(T_{\mathrm{enc}}^{\mathrm{base}}+T_{\mathrm{dec}}^{\mathrm{base}}\right)-T_{\mathrm{dec}}^{\mathrm{Para}}}
		\right\rceil,
		\label{eq:break_even}
	}
\end{equation}
provided that \(\left(T_{\mathrm{enc}}^{\mathrm{base}}+T_{\mathrm{dec}}^{\mathrm{base}}\right)>T_{\mathrm{dec}}^{\mathrm{Para}}\), where \(T_{\mathrm{offline}}^{\mathrm{Para}}\) is the one-time offline preparation time of ParaJSCC, and \(T_{\mathrm{enc}}^{\mathrm{base}}\), \(T_{\mathrm{dec}}^{\mathrm{base}}\), and \(T_{\mathrm{dec}}^{\mathrm{Para}}\) denote the baseline encoding time, baseline decoding time, and ParaJSCC decoding time, respectively. This metric quantifies the number of accesses required to amortize the offline cost of ParaJSCC under repeated serving; a smaller \(N^\star\) indicates that ParaJSCC becomes advantageous after fewer accesses.

\section{Network Instantiation of ParaJSCC}
\label{network}

\begin{figure}[t]
	\centering
	\includegraphics[width=0.9\linewidth]{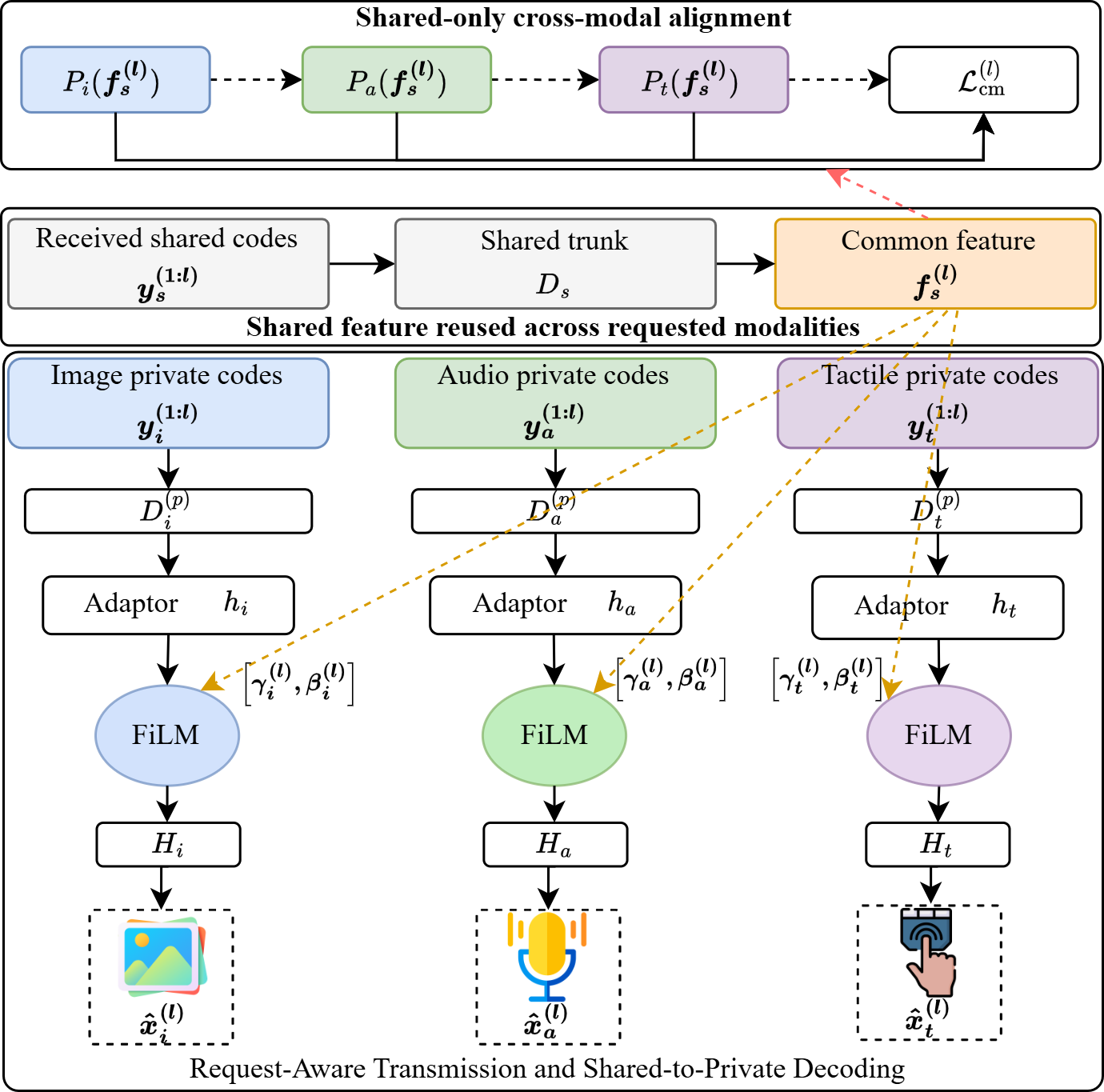}
	\caption{Architecture of the network instantiation of ParaJSCC. The received shared codes are decoded by the shared trunk \(D_s\) into a common feature \(\boldsymbol{f}_s^{(l)}\), which is reused across the requested modalities. For each modality \(m\), the private codes are decoded by \(D_m^{(p)}\), transformed by adaptor \(h_m\) into FiLM parameters, and passed to reconstruction head \(H_m\). During training, the shared feature is further projected by \(P_m\) to impose the shared-only cross-modal alignment loss \(\mathcal{L}_{\mathrm{cm}}^{(l)}\).}
	\label{fig:ParaJSCC_network}
\end{figure}

We instantiate the abstract formulation in Section~\ref{Methodology}, including the parameter package and decoder, as a concrete neural architecture. As shown in Fig.~\ref{fig:ParaJSCC_network}, the network instantiation of ParaJSCC represents each sample using progressive shared and modality-specific code tensors, and reconstructs the requested modalities through a shared-to-private decoder with cross-modal alignment.

\subsection{Progressive Shared-Private Code Tensors}

For each multimodal sample, the parameter groups defined in Section~\ref{Methodology} are instantiated as learnable shared and modality-specific code tensors arranged in a progressive pyramid:
\begin{equation}
	\boldsymbol{\Theta}_s^{(l)}\in\mathbb{R}^{C_s^{(l)}\times H_s^{(l)}\times W_s^{(l)}},\qquad
	\boldsymbol{\Theta}_m^{(l)}\in\mathbb{R}^{C_m^{(l)}\times H_m^{(l)}\times W_m^{(l)}},
\end{equation}
where \(l=1,\dots,L\) denotes the progressive level and \(m=1,\dots,M\) denotes the modality index. Lower levels provide coarse representations, while higher levels progressively refine them.

\subsection{Shared-to-Private Decoding}

Given a request \((\mathcal{T},l)\), let \(\boldsymbol{y}_s^{(1:l)}\) and \(\boldsymbol{y}_m^{(1:l)}\) denote the received shared and modality-specific codes. The decoder \(D_\phi\) follows a shared-to-private design.

The shared codes are first decoded into a common feature:
\begin{equation}
	\boldsymbol{f}_s^{(l)} = D_s\!\left(\boldsymbol{y}_s^{(1:l)}\right),
\end{equation}
where \(D_s(\cdot)\) is the shared decoder trunk.

For each requested modality \(m\in\mathcal{T}\), the private codes are decoded into a modulation feature:
\begin{equation}
	\boldsymbol{z}_m^{(l)} = D_m^{(p)}\!\left(\boldsymbol{y}_m^{(1:l)}\right),
\end{equation}
where \(D_m^{(p)}(\cdot)\) is the modality-specific private decoder. The modulation feature is then transformed by an adaptor \(h_m(\cdot)\) into FiLM~\cite{FiLM2018} parameters:
\begin{equation}
	[\boldsymbol{\gamma}_m^{(l)},\boldsymbol{\beta}_m^{(l)}] = h_m\!\left(\boldsymbol{z}_m^{(l)}\right),
\end{equation}
which modulate the shared feature as
\begin{equation}
	\tilde{\boldsymbol{f}}_m^{(l)} = \boldsymbol{\gamma}_m^{(l)} \odot \boldsymbol{f}_s^{(l)} + \boldsymbol{\beta}_m^{(l)}.
\end{equation}
The final reconstruction is produced by a modality-specific head:
\begin{equation}
	\hat{\boldsymbol{x}}_m^{(l)} = H_m\!\left(\tilde{\boldsymbol{f}}_m^{(l)}\right),\qquad m\in\mathcal{T}.
\end{equation}

Thus, the shared branch provides a common reconstruction basis, while the private branch injects modality-specific refinements through lightweight modulation.

\subsection{Shared-Only Alignment}

To encourage the shared parameters to encode cross-modal structure, the shared feature \(\boldsymbol{f}_s^{(l)}\) is projected into modality-aware embeddings:
\begin{equation}
	\boldsymbol{e}_m^{(s,l)} = P_m\!\left(\boldsymbol{f}_s^{(l)}\right),
\end{equation}
where \(P_m(\cdot)\) denotes the projection head for modality \(m\). These embeddings are aligned across the requested modalities using the cross-modal regularization defined in Section~\ref{Methodology}.

Accordingly, the global decoder is instantiated as
\begin{equation}
	D_\phi=\{D_s,\{D_m^{(p)},h_m,H_m,P_m\}_{m=1}^{M}\}.
\end{equation}

\section{Experiments}

\subsection{Experiment Settings}

\subsubsection{Experimental Setup and Hyperparameters}

All experiments were implemented in PyTorch on a workstation with an AMD EPYC 7F52 CPU (3.5~GHz) and an NVIDIA GeForce RTX 4090 GPU. We trained all models using AdamW with an initial learning rate of \(1 \times 10^{-5}\) and a StepLR scheduler with a step size of 100 epochs and a decay factor of 0.5. Unless otherwise specified, training was conducted for 100 epochs with a batch size of 8. Each multimodal sample consists of a \(256 \times 256\) image, an audio segment of length 1024, and a vibrotactile segment of length 576. The image modality is treated as two-dimensional data, whereas the audio and vibrotactile modalities are modeled as one-dimensional sequences.

For ParaJSCC, the request pair \((\mathcal{T}, l)\) is sampled during training from a factorized distribution \(p(\mathcal{T}, l)=p(\mathcal{T})p(l)\), covering all seven request types and six progressive levels. Unless otherwise specified, both \(p(\mathcal{T})\) and \(p(l)\) are chosen to be uniform, so that the training process does not depend on dataset-specific or application-specific request priors. The modality weights in Eq.~\ref{rm} are set to \(\lambda_m=0.2\), while the coefficients in Eq.~\ref{batagama} are fixed to \(\beta=0.1\) and \(\gamma=0.1\). ParaJSCC is trained with the proposed objective, whereas the baseline methods follow their original training objectives. All methods are evaluated under matched AWGN channel conditions. For non-JSCC codecs, the compressed bitstreams are transmitted over the same channel using digital modulation. Unless otherwise stated, all tabulated results are averaged over the two datasets.

For efficiency evaluation, we report receiver-side decoding latency and repeated-serving cost. Decoding latency measures only the reconstruction time after the transmitted symbols, bitstreams, or parameter subsets have been received. Repeated-serving cost is defined as the amortized computation cost over multiple accesses: for ParaJSCC, it consists of one-time offline parameter preparation plus per-access decoding, whereas for conventional baselines, it consists of per-access encoding and decoding. Because existing single-modality baselines do not support request-aware repeated multimodal serving, multimodal requests are emulated by sequentially executing the corresponding modality-specific pipelines.

\subsubsection{Dataset}

We evaluate the proposed method on two public multimodal surface-interaction datasets: LMT-108~\cite{LMT108-2016} and BFTS~\cite{BFTS2025}. These datasets provide complementary visual, auditory, and vibrotactile signals. LMT-108 contains 108 textured surfaces with paired image, audio, and tactile recordings collected during tool-mediated freehand exploration. BFTS was collected during bare-finger exploration and includes 500 trials from 10 participants interacting with 10 representative textures, together with synchronized image, audio, and friction-induced vibration signals. For both datasets, samples are randomly split into training and validation sets with a ratio of 7:3 while preserving the material-class distribution. All reported results are averaged over the validation sets.

\subsubsection{Comparison Algorithms}

We compare ParaJSCC against three categories of baselines: modality-specialized methods, unified online multimodal methods, and reusable but unstructured representations, since no existing method fully matches the proposed repeated-access multimodal serving protocol.

\textbf{Modality-specialized baselines.} For image transmission, we consider ADJSCC~\cite{ADJSCC2022}, JSCCformer~\cite{JSCCformer2024}, CDDM~\cite{CDDM2024}, and SwinJSCC~\cite{Swinjscc2025}. For audio transmission, we use DeepSC-S~\cite{DeepSC-S2021}, DNN~\cite{DNN2023}, and DNN2~\cite{DNN22025} as speech-oriented neural acoustic transmission baselines, since dedicated JSCC methods for the non-speech surface-interaction audio signals in our datasets remain limited. These methods therefore serve as the closest available baselines for one-dimensional acoustic signal communication under the same AWGN setting. For vibrotactile transmission, we compare PVC-SLP~\cite{PVC-SLP2020}, VC-PWQ~\cite{VC-PWQ2021}, RNVC~\cite{RNVC2022}, and VTSC~\cite{VTSC2025}. Since these tactile codecs are separation-based, their compressed outputs are further transmitted over the same AWGN channel using 16-QAM modulation. Because these baselines do not support repeated multimodal serving, multimodal requests are handled by sequentially executing the corresponding modality-specific pipelines.

\textbf{Unified online multimodal baselines.} We include U-DeepSC~\cite{U-DeepSC2024} as a representative unified multimodal baseline. We further implement a Unified Online JSCC (UniJSCC) baseline, which adopts a unified multimodal encoder-decoder with a shared latent representation and modality-specific heads, but re-encodes the raw multimodal source for every request. This baseline isolates the gain of reusable offline serving from the gain of using a unified multimodal architecture alone.

\textbf{Reusable representation baseline.} We also implement a Flat Reusable JSCC (FlatJSCC) baseline. Similar to ParaJSCC, it performs offline per-instance optimization and stores a quantized representation for repeated access. However, unlike ParaJSCC, it uses a single flat parameter package without shared-private decomposition, progressive hierarchy, or request-aware subset selection. As a result, the entire stored package must be transmitted for each request.

\subsubsection{Evaluation Metrics}

We use standard reconstruction metrics for the three modalities. Specifically, PSNR and SSIM are reported for image reconstruction, SI-SDR~\cite{le2019sdr} for audio reconstruction, and ST-SIM~\cite{STSIM2019} for vibrotactile reconstruction.

\begin{figure*}[t]
	\centering
	\includegraphics[width=\linewidth]{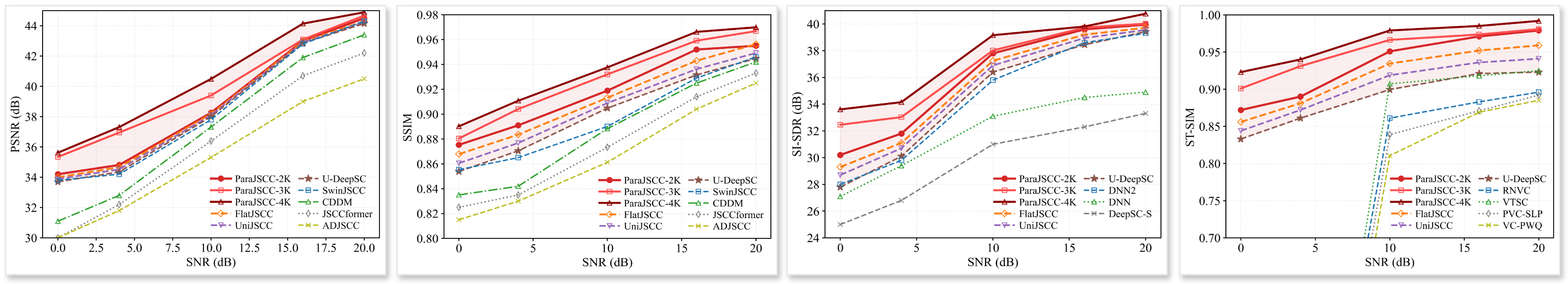}
	\caption{Performance under varying SNR at a fixed bandwidth ratio on the LMT-108 dataset, including ParaJSCC with 2K, 3K, and 4K offline optimization iterations.}
	
	\label{fig:lmt108_snr}
\end{figure*}

\begin{figure*}[t]
	\centering
	\includegraphics[width=\linewidth]{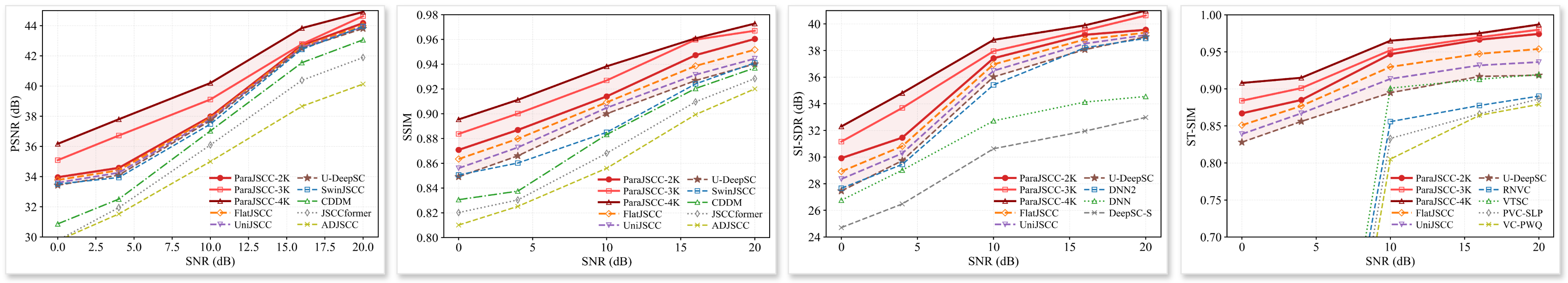}
	\caption{Performance under varying SNR at a fixed bandwidth ratio on the BFTS dataset, including ParaJSCC with 2K, 3K, and 4K offline optimization iterations.}
	\label{fig:bfts_snr}
\end{figure*}

\subsection{Performance under Varying SNR at a Fixed Bandwidth Ratio}

We evaluate ParaJSCC under varying channel conditions at a fixed bandwidth ratio \(R=0.4406\) using the LMT-108 and BFTS datasets. Comparisons include modality-specific baselines, U-DeepSC, UniJSCC, and FlatJSCC, all operating under the same transmission budget, alongside ParaJSCC with 2K, 3K, and 4K offline optimization iterations. As shown in Figs.~\ref{fig:lmt108_snr} and \ref{fig:bfts_snr}, ParaJSCC achieves the strongest overall performance across image, audio, and vibrotactile modalities on both datasets. For image and audio reconstruction, increasing the offline optimization depth consistently improves performance, particularly at low and medium SNRs. For vibrotactile reconstruction, ParaJSCC attains the highest ST-SIM scores, while several conventional baselines degrade sharply at low SNR due to the cliff effect associated with separate source coding under fixed 16-QAM modulation.

Among the unified baselines, UniJSCC outperforms U-DeepSC, while FlatJSCC further improves upon UniJSCC, demonstrating the benefits of reusable offline representations. ParaJSCC surpasses FlatJSCC, particularly at low and medium SNRs, validating the advantage of the proposed progressive shared-private parameterization. Although deeper offline optimization remains beneficial, its gains diminish at high SNRs.

\subsection{Online Efficiency and Repeated-Access Benefit}

\begin{table}[t]
	\centering
	\setlength{\tabcolsep}{2.5pt}
	\renewcommand{\arraystretch}{0.85}
	\caption{Receiver-side online decoding latency under different request types. Latency is measured only for reconstruction after the transmitted symbols, bitstreams, or parameter subsets have been received.}	
	\label{tab:latency}
	\begin{tabular}{lccccc}
		\toprule
		\multirow{2}{*}{Request} & \multicolumn{5}{c}{Latency (ms)} \\
		\cmidrule(lr){2-6}
		& Baseline & U-DeepSC & UniJSCC & FlatJSCC & ParaJSCC \\
		\midrule
		Img     & 17.18 & 13.65 & 14.42 & 5.12  & 4.34 \\
		Aud     & 8.09  & 5.37  & 5.94  & 2.58  & 2.16 \\
		Tac     & 16.83 & 5.01  & 5.63  & 2.49  & 2.11 \\
		Img+Aud & 25.95 & 18.32 & 19.41 & 7.54  & 6.71 \\
		Img+Tac & 35.33 & 17.03 & 18.65 & 7.31  & 6.53 \\
		Aud+Tac & 25.82 & 9.64  & 10.82 & 4.86  & 4.32 \\
		All     & 43.96 & 24.31 & 26.48 & 12.46 & 11.21 \\
		\bottomrule
	\end{tabular}
\end{table}

We evaluate ParaJSCC in terms of online request-time latency and long-term repeated-access efficiency. Table~\ref{tab:latency} reports the receiver-side decoding latency after the transmitted symbols, bitstreams, or parameter subsets have been received, excluding source encoding and offline preparation. For fairness, both ParaJSCC and FlatJSCC use 2,000 offline optimization iterations. All methods are evaluated on the same hardware with a batch size of 1, and latency is averaged over 1,000 runs following 100 warm-up runs.

ParaJSCC achieves the lowest decoding latency across all request types. FlatJSCC consistently outperforms the online baselines, demonstrating that reusable offline representations reduce request-time overhead. ParaJSCC further improves upon FlatJSCC by decoding only the shared and modality-specific subsets required for the current request, rather than the entire reusable package. This advantage is especially pronounced for multimodal requests, where the performance gap relative to U-DeepSC and UniJSCC increases with the number of requested modalities.

For long-term efficiency, we use the break-even access count \(N^\star\) as defined in Section~\ref{Methodology}. ParaJSCC and FlatJSCC incur a one-time offline preparation cost of 20.0 seconds per sample under the 2K-iteration setting, whereas the composed modality-specialized baseline performs encoding and decoding for every access. Compared to this baseline, the image-only per-access cost decreases from 27.32 ms to 4.34 ms, resulting in \(N^\star = 871\). For full multimodal requests, the cost decreases from 61.48 ms to 11.21 ms, yielding \(N^\star = 398\). The lower break-even count for the full request indicates that the benefits of reusable serving become more pronounced as request complexity increases.

\subsection{Flexible Serving under Heterogeneous Requests}

\begin{table}[t]
	\centering
	\setlength{\tabcolsep}{2.5pt}
	\renewcommand{\arraystretch}{0.85}
	\caption{Request-aware reconstruction under different modality requests (SNR = 10 dB).}
	\label{tab:request_selective}
	\begin{tabular}{lcccccc}
		\toprule
		Request type & Rate & Latency (ms) & PSNR & SSIM & SI-SDR & ST-SIM \\
		\midrule
		Img     & 0.2302 & 4.34  & 38.26 & 0.919 & --    & --    \\
		Aud     & 0.2154 & 2.16  & --    & --    & 37.50 & --    \\
		Tac     & 0.2150 & 2.11  & --    & --    & --    & 0.912 \\
		Img+Aud & 0.3356 & 6.71  & 38.03 & 0.907 & 37.60 & --    \\
		Img+Tac & 0.3352 & 6.53  & 38.05 & 0.910 & --    & 0.915 \\
		Aud+Tac & 0.3204 & 4.32  & --    & --    & 37.65 & 0.934 \\
		All     & 0.4406 & 11.21 & 38.18 & 0.921 & 37.72 & 0.938 \\
		\bottomrule
	\end{tabular}
\end{table}

\begin{table}[t]
	\centering
	\setlength{\tabcolsep}{2.5pt}
	\renewcommand{\arraystretch}{0.85}
	\caption{Progressive reconstruction results under different request types.}
	\label{tab:prog_all}
	\begin{tabular}{lcccccc}
		\toprule
		Request & Level & Rate & PSNR & SSIM & SI-SDR & ST-SIM \\
		\midrule
		\multirow{6}{*}{Img}
		& 1 & 0.0814 & 30.85 & 0.865 & -- & -- \\
		& 2 & 0.1197 & 33.72 & 0.892 & -- & -- \\
		& 3 & 0.1568 & 35.82 & 0.905 & -- & -- \\
		& 4 & 0.1895 & 36.98 & 0.912 & -- & -- \\
		& 5 & 0.2124 & 37.78 & 0.916 & -- & -- \\
		& 6 & 0.2302 & 38.26 & 0.919 & -- & -- \\
		\midrule
		\multirow{6}{*}{Aud+Tac}
		& 1 & 0.1137 & -- & -- & 25.20 & 0.845 \\
		& 2 & 0.1683 & -- & -- & 30.80 & 0.885 \\
		& 3 & 0.2196 & -- & -- & 33.90 & 0.905 \\
		& 4 & 0.2639 & -- & -- & 35.60 & 0.918 \\
		& 5 & 0.2958 & -- & -- & 36.80 & 0.928 \\
		& 6 & 0.3204 & -- & -- & 37.65 & 0.934 \\
		\midrule
		\multirow{6}{*}{All}
		& 1 & 0.1542 & 30.20 & 0.858 & 24.80 & 0.835 \\
		& 2 & 0.2287 & 33.10 & 0.888 & 30.20 & 0.875 \\
		& 3 & 0.2984 & 35.10 & 0.902 & 33.50 & 0.898 \\
		& 4 & 0.3573 & 36.40 & 0.910 & 35.40 & 0.915 \\
		& 5 & 0.4041 & 37.30 & 0.915 & 36.70 & 0.926 \\
		& 6 & 0.4406 & 38.18 & 0.921 & 37.72 & 0.938 \\
		\bottomrule
	\end{tabular}
\end{table}

We evaluate the flexible serving capability of ParaJSCC for heterogeneous requests from two complementary perspectives: request-aware selective reconstruction and progressive quality adaptation. Since the compared baselines do not support structured subset transmission or progressive parameter hierarchies, this experiment evaluates ParaJSCC independently.

Table~\ref{tab:request_selective} demonstrates that ParaJSCC supports all seven non-empty modality subsets by transmitting only the shared parameters and the modality-specific subsets required for each request. Compared to the full multimodal request (\(R=0.4406\)), single-modality requests require rates of only \(0.2150\) to \(0.2302\), reducing the transmission rate by approximately 48\% to 51\% and lowering decoding latency from 11.21 ms to between 2.11 and 4.34 ms. Meanwhile, reconstruction quality remains stable: the image-only request achieves 38.26 dB PSNR and 0.919 SSIM, compared with 38.18 dB and 0.921 for the full request, while audio SI-SDR and tactile ST-SIM vary narrowly from 37.50 to 37.72 dB and from 0.912 to 0.938, respectively.

Table~\ref{tab:prog_all} further validates progressive reconstruction for image-only, audio+tactile, and full multimodal requests across levels \(l=1\) to \(l=6\). Each reconstruction utilizes only the parameter groups from levels \(1{:}l\), with quality improving monotonically as the transmission rate increases. Image PSNR rises from 30.85 to 38.26 dB, while audio+tactile SI-SDR and ST-SIM increase from 25.20 dB and 0.845 to 37.65 dB and 0.934, respectively. Full multimodal reconstruction exhibits the same trend. These results confirm that ParaJSCC reduces communication costs through request-aware subset transmission while enabling coarse-to-fine reconstruction without requiring request-time re-encoding or full-package transmission. More importantly, the progressive shared-private representation allows a single stored multimodal asset to adapt to dynamically changing user requirements and bandwidth conditions without additional model redesign. This property makes ParaJSCC suitable for long-term serving scenarios involving persistent multimodal content, such as multimodal repositories and digital twins.

\subsection{Ablation Study on Core Designs}

\begin{table}[t]
	\centering
	\setlength{\tabcolsep}{2.2pt}
	\renewcommand{\arraystretch}{0.85}
	\caption{Ablation results under the full multimodal request (SNR = 10 dB).}
	\label{tab:ablation_full}
	\begin{tabular}{lcccc}
		\toprule
		Variant & Latency (ms) & PSNR & SI-SDR & ST-SIM \\
		\midrule
		Full ParaJSCC & 11.21 & 38.18 & 37.72 & 0.938 \\
		w/o Shared-Private & 10.09 & 36.95 & 36.20 & 0.915 \\
		w/o Progressive & 8.34 & 37.30 & 36.85 & 0.925 \\
		w/o Alignment & 10.93 & 37.55 & 37.05 & 0.928 \\
		w/o Request-Aware Selection & 11.43 & 38.05 & 37.60 & 0.936 \\
		\bottomrule
	\end{tabular}
\end{table}

We conduct an ablation study on four core components of ParaJSCC: shared-private parameterization, progressive hierarchy, shared-only cross-modal alignment, and request-aware subset selection. The full model is compared with four corresponding ablated variants: w/o Shared-Private, w/o Progressive, w/o Alignment, and w/o Request-Aware Selection. Here, w/o Shared-Private is an internal ablation within the ParaJSCC framework, whereas FlatJSCC is treated separately as a standalone baseline in the previous experiments.

As shown in Table~\ref{tab:ablation_full}, removing any component degrades performance, confirming that each design element contributes to the effectiveness of ParaJSCC. Among them, shared-private parameterization is the most critical, as its removal causes the largest quality drop across all three modalities. The progressive hierarchy also provides substantial gains, indicating that it improves not only scalability but also representation quality under a fixed transmission budget. Removing the alignment loss leads to a smaller but consistent degradation, suggesting that it helps the shared branch capture more modality-invariant information. 
In contrast, removing request-aware subset selection has only a minor effect under the full multimodal request, where all parameters are transmitted; its main benefit lies in improving efficiency under partial requests. Overall, these results verify the effectiveness of each component.

\section{Conclusion}
This paper proposes ParaJSCC, a parameterized multimodal JSCC framework for repeated-access multimodal communication. By converting each multimodal sample into a reusable parameter package through offline per-instance optimization, ParaJSCC replaces repeated request-time encoding with request-aware subset transmission and lightweight reconstruction. Its progressive shared-private parameterization further enables cross-modal sharing, modality-selective transmission, and coarse-to-fine quality adaptation under varying bandwidth constraints. Experiments involving image, audio, and vibrotactile communication demonstrate that ParaJSCC delivers high reconstruction quality, robust performance under noisy channels, and clear advantages in selective serving, progressive reconstruction, and repeated-access efficiency. Overall, these results suggest that ParaJSCC provides an effective foundation for serving persistent multimodal assets in cloud-edge multimodal systems, including multimodal repositories and digital twins.

\bodyrefbreak

\bibliographystyle{ACM-Reference-Format}
\bibliography{reference}

\end{document}